\documentclass[aps,prb,twocolumn,showpacs,amsmath,superscriptaddress,floatfix]{revtex4}

\usepackage{amsmath}
\usepackage{amsfonts}
\usepackage{bm}
\usepackage{epsfig}
\usepackage{graphicx}
\usepackage{graphics}
\usepackage{url}
\usepackage[english]{babel}
\usepackage{dcolumn}
\usepackage{color}

\begin{document}

\title{Energy transport and coherence properties of acoustic phonons generated by
 optical excitation of a quantum dot}

\author{D.~Wigger}
\affiliation{Institut f\"ur Festk\"orpertheorie, Universit\"at M\"unster, Wilhelm-Klemm-Str.~10, 48149 M\"unster,
Germany}

\author{S.~L\"uker}
\affiliation{Institut f\"ur Festk\"orpertheorie, Universit\"at M\"unster, Wilhelm-Klemm-Str.~10, 48149 M\"unster,
Germany}

\author{D.~E.~Reiter}
\affiliation{Institut f\"ur Festk\"orpertheorie, Universit\"at M\"unster, Wilhelm-Klemm-Str.~10, 48149 M\"unster,
Germany}

\author{V.~M.~Axt}
\affiliation{Theoretische Physik III, Universit\"at Bayreuth, 95440
Bayreuth,
Germany}

\author{P.~Machnikowski}
\affiliation{Institute of Physics, Wroc{\l}aw University of Technology,
50-370 Wroc{\l}aw, Poland}

\author{T.~Kuhn}
\affiliation{Institut f\"ur Festk\"orpertheorie, Universit\"at M\"unster, Wilhelm-Klemm-Str.~10, 48149 M\"unster,
Germany}

\date{\today}

\begin{abstract}
The energy transport of acoustic phonons generated by the optical excitation
of a quantum dot as well as the coherence properties of these phonons are
studied theoretically both for the case of a pulsed excitation and for a
continuous wave (cw) excitation switched on instantaneously. For a pulsed
excitation, depending on pulse area and pulse duration, a finite number of
phonon wave packets is emitted, while for the case of a cw excitation a
sequence of wave packets with decreasing amplitude is generated after the
excitation has been switched on. We show that the energy flow associated with
the generated phonons is partly related to coherent phonon oscillations and
partly to incoherent phonon emission. The efficiency of the energy transfer
to the phonons and the details of the energy flow depend strongly and in a
non-monotonic way on the Rabi frequency exhibiting a resonance behavior.
However, in the case of cw excitation it turns out that the total energy
transferred to the phonons is directly linked in a monotonic way to the Rabi
frequency.
\end{abstract}


\maketitle

\section{Introduction}

The interaction between electrons and phonons plays a vital role in the
optical control of semiconductor quantum dots (QDs). Besides its often
decisive role for the dephasing and decoherence of electronic excitations in
QDs, there has been an increasing interest in the properties of the phonons
themselves, which are generated by optical excitation of the QD, and in the
use of phonons for an active control of the optical properties of such
nanostructures. Propagating strain waves, i.e., coherent phonon pulses, which
travel through a QD or quantum well structure give rise to a dynamical shift
of the optical transition frequency \cite{akimov2006ult, scherbakov2007chi,
huneke2008imp, gotoh2013mod} which may even lead to an ultrafast switching
into the lasing regime.\cite{bruggemann2011las} Coherent acoustic phonons can
also be used to sweep the frequency of optical cavities across the resonance
frequencies of QDs.\cite{blattmann2014ent} The generation of coherent phonons
\cite{cho1990sub,
merlin1997gen,misochko2001coh,devos2007str,papenkort2010res} and phonons in
non-classical states such as squeezed phonons has been of considerable
interest.\cite{hu1996squ,misochko2000pha,sauer2010lat,hussain2010abs} An
optically driven QD has been proposed for the realization of a phonon laser
when placed in an acoustic cavity\cite{kabuss2012opt} or it could act as a
heat pump when driven non-resonantly.\cite{gauger2010hea} Especially in
systems that reach the quantum-classical boundary the understanding of the
electron-phonon coupling mechanism and the characteristic phonon properties
on a microscopical level is crucial.
\cite{wilson2004las,rundquist2011off,yeo2013str}

The impact of the electron-phonon coupling on optically induced exciton
dynamics in QDs has been extensively studied in theory and experiments.
Besides giving rise to a finite linewidth and a characteristic non-Lorentzian
line shape of luminescence, \cite{besombes2001aco}
absorption,\cite{krummheuer2002the,stock2011aco} or four-wave-mixing
spectra,\cite{borri2001ult,vagov2004non} the coupling to phonons causes a
damping of Rabi oscillations of the electronic system \cite{kamada2001exc,
zrenner2002coh, forstner2003pho, machnikowski2004res, borri2005exc,
krugel2005the, vagov2007non, mccutcheon2010qua, ramsay2010dam, ramsay2010pho}
and it limits the achievable exciton population inversion by chirped laser
pulses via the adiabatic rapid passage.\cite{simon2011rob, wu2011pop,
luker2012inf, debnath2012chi, eastham2013lin} The efficiency of the
electron-phonon coupling in the presence of a strong optical driving depends
crucially and in a non-monotonic way on the laser power leading to the
reappearance of Rabi rotations \cite{vagov2007non,ramsay2010pho} at
sufficiently high pulse areas.

The main focus of the present paper is on the detailed analysis of the
characteristic properties and the spatio-temporal dynamics of phonons
generated by optical excitation of a QD, in particular on the degree of
coherence and the energy transfer to the phonons. Furthermore we will show
that studying the phonon dynamics will provide additional insight in the
non-monotonic behavior of Rabi rotations as a function of the light
intensity.

The QD exciton is coupled to both acoustic and optical phonons. The optical
phonons, due to their in general negligible dispersion in the range of wave
vectors which are coupled to the exciton, give rise to an oscillatory
contribution in the optical polarization and sharp replicas in the optical
spectra.\cite{krummheuer2002the,stock2011aco} Because of their vanishing
group velocity, the optical phonons generated by this coupling remain
confined to the QD region.\cite{sauer2010lat, reiter2011gen} As long as
anharmonic processes can be neglected they do not lead to an irreversible
decay of the polarization. Acoustic phonons, on the other hand, provide a
continuum of states to which the QD exciton is coupled and, indeed, the
coupling to acoustic phonons has been shown to be typically the main source
of decoherence in optically driven QDs.\cite{vagov2004non,ramsay2010pho}
Therefore, in this paper we will concentrate on the coupling to acoustic
phonons.

For an excitation with ultrafast laser pulses, which can be approximated by
$\delta$-functions in time, the coupled exciton-phonon dynamics can be
calculated analytically.\cite{vagov2002ele, wigger2013flu} For this case, it
has been found that by the impulsive excitation a lattice deformation in the
region of the QD is created forming the acoustic polaron, which is
accompanied by the emission of a phononic wave packet.\cite{vagov2002ele,
jacak2003coh, krummheuer2005pur, machnikowski2007qua, wigger2013flu} By
excitation with a tailored series of pulses the fluctuations of the phonons
can be modified such that squeezed phonon wave packets can be
generated.\cite{wigger2013flu} In the case of excitation by sufficiently
slowly varying light fields, on the other hand, the polaron builds up
adiabatically and no phonon wave packet is emitted.\cite{machnikowski2007qua}

In this paper we will study the coupled exciton-phonon dynamics in the case
of excitation by pulses with a finite pulse width as well as by a continuous
light field with fixed amplitude which is instantaneously switched on. While
in the absence of a light field the electron-phonon interaction in a QD is of
pure-dephasing type, i.e., there are no real transitions between different
electronic states, in the presence of a light field the interaction gives
rise to transitions between the dressed states and therefore at sufficiently
low temperatures typically to a relaxation into the lower dressed state. The
two cases studied here thus represent two different aspects of
non-equilibrium dynamical systems: For pulsed excitation a relaxation channel
is switched on for a certain period of time and then the system returns to
the pure-dephasing type. For the continuous excitation, due to the
instantaneous switching on, the system is prepared in a non-equilibrium
distribution over the dressed states and then it relaxes towards a new
stationary state.

For both excitation conditions we will analyze quantitatively the energy
transfer from the exciton to the phonon system and the energy flow associated
with the emission of phonon wave packets. We will distinguish between the
energy transport by coherent phonons, i.e., phonons with a non-vanishing
expectation value of the displacement corresponding to a classical strain
wave, and the energy flow associated with the emission of incoherent phonons,
which are characterized by a vanishing mean displacement. In addition to a
detailed characterization of the traveling phonon waves, the direct
comparison of the time evolution of excitonic and phononic variables for
varying pulse durations and light intensities will provide a common
perspective on the close link between exciton dephasing and irreversibility
introduced by the energy flow away from the QD by the traveling acoustic
phonons and on the resonant nature of the exciton-phonon coupling in these
structures. In particular we will show that although the efficiency of
electron-phonon coupling is a strongly non-monotonic function of the light
intensity, in the case of continuous excitation the total energy carried away
by the phonons is monotonically increasing with the light intensity.

The paper is organized as follows: After the introduction, in
Sec.~\ref{sec:model} we summarize the theoretical background and the methods
which have been used to calculate the coupled exciton-phonon dynamics. In
Sec.~\ref{sec:pulse} we begin the discussion of the results by analyzing the
phonon dynamics and the corresponding energy transport resulting from a
pulsed excitation of the QD. Section~\ref{sec:cw} is then devoted to the
coupled exciton-phonon dynamics and the energy transport in the case of a
continuous wave excitation switched on instantaneously. Finally, in
Sec.~\ref{sec:conclusions} we will summarize our results and provide some
concluding remarks.

\section{Model system}\label{sec:model}

We consider a QD driven by a circularly polarized light field. The QD is
taken to be in the strong confinement limit such that the electronic degrees
of freedom can be reduced to a two-level system consisting of the ground
state $|g\rangle$ and the single exciton state $|x\rangle$ with the angular
momentum determined by the polarization of the light. The electronic system
is coupled to bulk acoustic phonons via the pure dephasing mechanism, i.e.,
phonon-induced transitions to other electronic states are neglected because
the energy separations are assumed to be much larger than the phonon
energies. The Hamiltonian of the system then reads
\begin{eqnarray}
\hat{H} &=& \hbar\omega_x |x\rangle\langle x|-
\mathbf{M}\cdot\mathbf{E}^{(+)}|x\rangle \langle g| -
\mathbf{M}^\ast\cdot\mathbf{E}^{(-)}|g\rangle \langle x| \notag\\
&&+\sum_{\mathbf{q}}\hbar\omega_{\mathbf{q}}\hat{b}_{\mathbf{q}}^\dag
\hat{b}_{\mathbf{q}}^{}
+\sum_{\mathbf{q}}\hbar\left(g_{\mathbf{q}}\hat{b}_{\mathbf{q}} +
g_{\mathbf{q}}^\ast \hat{b}_{\mathbf{q}}^\dag\right)|x\rangle\langle x|, \label{eq:H}
\end{eqnarray}
where $\hbar\omega_x$ denotes the exciton energy, $g_{\mathbf{q}}$ is the
carrier-phonon coupling matrix element, $\hat{b}_{\mathbf{q}}^\dag$
($\hat{b}_{\mathbf{q}}$) are the creation (annihilation) operators for a
phonon with wave vector $\mathbf{q}$ and energy $\hbar\omega_{\mathbf{q}}$.
$\mathbf{E}^{(\pm)}$ denotes the positive (negative) frequency component of
the driving light field, which is coupled to the carriers in the usual
rotating wave and dipole approximation via the dipole matrix element
$\mathbf{M}$. We set
\begin{equation}
\mathbf{M}\cdot\mathbf{E}^{(+)} = \frac{\hbar}{2}\Omega(t) e^{-i
\omega_{\textrm{L}} t},\label{eq:rabi}
\end{equation}
where $\Omega(t)$ denotes the instantaneous Rabi frequency and
$\omega_{\textrm{L}}$ is the central frequency of the light field. In all the
calculations discussed here we assume a resonant excitation, i.e., the light
frequency agrees with the polaron-shifted exciton energy according to
$\omega_{\textrm{L}} = \omega_x - \omega_{\textrm{pol}}$
with\cite{krummheuer2002the}
\begin{equation} \omega_{\textrm{pol}}=\sum_{\mathbf{q}} \frac{\left|
g_{\mathbf{q}} \right|^2}{\omega_{\mathbf{q}}}. \label{eq:polaron}
\end{equation}

For simplicity we consider a spherical QD with a harmonic confinement
potential and restrict ourselves to the coupling to longitudinal acoustic
(LA) phonons via the deformation potential interaction. Unless there is a
strong spatial separation of electron and hole states, this is typically the
dominant mechanism for phonon-induced dephasing.\cite{vagov2004non} A
non-spherical QD shape would introduce an angular dependence in the phonon
coupling and thus also in the emitted wave packet,\cite{krummheuer2005pur}
however the overall phonon emission dynamics is unchanged. We assume
bulk-like LA phonons with a linear dispersion relation given by
$\omega_{\mathbf{q}}= c_l q$, $c_l$ being the longitudinal sound velocity.
Due to the harmonic confinement potential, the deformation potential coupling
matrix element can be written in a closed form; it reads\cite{vagov2002ele}
\begin{equation}
g_{\mathbf{q}} = \sqrt{\frac{1}{2\rho \hbar V
\omega_{\mathbf{q}}}}q\left(D_{\textrm{e}} e^{-(q a_{\textrm{e}}/2)^2} - D_{\textrm{h}}
e^{-(q a_{\textrm{h}}/2)^2}\right),
\end{equation}
where $V$ is a normalization volume, $\rho$ is the material density,
$D_{\textrm{e(h)}}$ are the deformation potentials for electrons (holes), and
$a_{\textrm{e(h)}}$ are the localization lengths of the electron (hole) wave
functions. Assuming identical confinement potentials for electrons and holes,
the localization lengths of electron and hole are related by the ratio of
electron and hole mass according to
$a_{\textrm{h}}/a_{\textrm{e}}=(m_{\textrm{e}}/m_{\textrm{h}})^{1/4} \approx
0.87$.\cite{krummheuer2002the}

\begin{figure}
\includegraphics[width=\columnwidth]{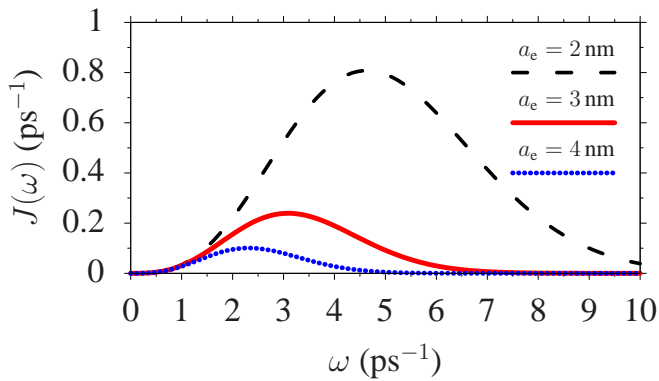}
\caption{(Color online) Phonon spectral density $J(\omega)$ for quantum dot sizes
$a_{\textrm{e}}=2\,{\textrm{nm}}, 3\,{\textrm{nm}}, {\textrm{and}}\ 4\,{\textrm{nm}}$.
Calculations in this
paper have been performed for $a_{\textrm{e}}=3\,{\textrm{nm}}$.}
\label{fig:spectral}
\end{figure}

The frequency-dependent strength of the coupling to the phonons is described
by the phonon spectral density
\begin{equation}
J(\omega) = \sum_{\mathbf{q}} \left| g_{\mathbf{q}}\right|^2 \delta (\omega-\omega_{\mathbf{q}}),
\label{eq:spectral}
\end{equation}
which can directly be calculated from the coupling matrix
element.\cite{jacak2003coh, krugel2005the, axt2005red}
Figure~\ref{fig:spectral} shows this spectral density for QDs of three
different sizes. The spectral density has a finite bandwidth; both the
bandwidth and the maximum depend on the size expressed in terms of the
electron localization $a_{\textrm{e}}$ of the dot. For decreasing QD size,
the peak position of the spectral density shifts to higher phonon frequencies
and it increases in strength. In the following we will focus on a QD with an
electron localization length $a_{\textrm{e}}=3\,{\textrm{nm}}$. In this case
the spectral density has a maximum at approximately $\omega_{\textrm{ph}} =
3\,{\textrm{ps}}^{-1}$, therefore the modes with this frequency couple most
strongly to the exciton. The oscillation period of the most strongly coupled
phonon modes is then $T_{\textrm{ph}}=2\pi/\omega_{\textrm{ph}}\approx2$~ps.
The coupling is essentially restricted to phonons with frequencies between 1
and 6~ps$^{-1}$.

In this paper we are mainly interested in the energy which is transferred to
the acoustic phonon system and which is carried away from the QD region by
these phonons. Outside the QD region there is no energy transferred to or
from the phonons leading to energy conservation in the phonon system. From
elasticity theory it is known that, similar to the case of electrodynamics,
this energy conservation is reflected in an \emph{acoustic Poynting
theorem}\cite{auld1973aco}
\begin{equation}
\frac{\partial w_{\textrm{LA}}({\mathbf{r}},t)}{\partial t}
+ \textrm{div} \left( {\mathbf{S}}_{\textrm{LA}}({\mathbf{r}},t) \right) = 0
\end{equation}
with the acoustic energy density $w_{\textrm{LA}}({\mathbf{r}},t)$ and the
acoustic Poynting vector ${\mathbf{S}}_{\textrm{LA}}({\mathbf{r}},t)$. In our
case, where only LA phonons are created, the acoustic Poynting theorem can be
obtained from the wave equation for the LA phonons,
\begin{equation}
\frac{\partial^2 {\mathbf{u}}}{\partial t^2}- c_l^2 \Delta {\mathbf{u}} =0 ,
\end{equation}
which leads to the energy density
\begin{equation}
w_{\textrm{LA}}({\mathbf{r}},t)=\frac{\rho}{2}
\left(\frac{\partial {\mathbf{u}}({\mathbf{r}},t)}{\partial t}\right)^2
+ \frac{1}{2}\rho c_l^2 \sum_{i,j} \left(\frac{\partial u_i({\mathbf{r}},t)}{\partial x_j}\right)^2
\label{eq:energy-density-LA}
\end{equation}
and the $i$-th component of the associated acoustic Poynting vector
${\mathbf{S}}_{\textrm{LA}}$
\begin{equation}
\left( S_{\textrm{LA}} \right)_i = -\rho c_l^2 \sum_{j}
\frac{\partial u_j({\mathbf{r}},t)}{\partial t} \frac{\partial u_j({\mathbf{r}},t)}{\partial x_i} .
\label{eq:poynting-LA}
\end{equation}
We want to remark that Eqs.~\eqref{eq:energy-density-LA} and
\eqref{eq:poynting-LA} differ from the general formulas obtained from
elasticity theory\cite{auld1973aco} due to the restriction to an isotropic
medium with only LA phonons.

In a quantum mechanical treatment, as is necessary here, the lattice
displacement ${\mathbf{u}}$ is replaced by the corresponding operator
$\hat{{\mathbf{u}}}$ which, when expressed in terms of the phonon creation
and annihilation operators for the LA phonons, reads
\begin{equation}
\hat{ {\mathbf{u}} }({\mathbf{r}}) = - i \sum_{\mathbf{q}} \sqrt{\frac{\hbar}{2\rho V
\omega_{\mathbf{q}}}}\left(\hat{ b}_{\mathbf{q}} e^{i{\mathbf{q}}\cdot{\mathbf{r}}} -
\hat{ b}_{\mathbf{q}}^\dag e^{-i{\mathbf{q}}\cdot{\mathbf{r}}}\right)\frac{{\mathbf{q}}}{q} .
\label{eq:u}
\end{equation}
Accordingly, the temporal derivative $\partial \mathbf{u} / \partial t$ is
replaced by $\hat{\bm{\pi}} / \rho $ with the conjugate momentum operator
\begin{equation}
\hat{ {\bm{\pi}} }({\mathbf{r}}) = -  \sum_{\mathbf{q}} \sqrt{\frac{\rho\hbar
\omega_{\mathbf{q}}}{2 V}}\left(\hat{ b}_{\mathbf{q}} e^{i{\mathbf{q}}\cdot{\mathbf{r}}} +
\hat{ b}_{\mathbf{q}}^\dag e^{-i{\mathbf{q}}\cdot{\mathbf{r}}}\right)\frac{{\mathbf{q}}}{q} .
\label{eq:pi}
\end{equation}
Then also the acoustic energy density and the acoustic Poynting vector become
operators $\hat{w}_{\textrm{LA}}$ and $\hat{\mathbf{S}}_{\textrm{LA}}$,
respectively, which can be expressed in terms of the phonon creation and
annihilation operators. Using the standard Bose commutation relations for
these operators and integrating the energy density operator over the whole
space, the free phonon Hamiltonian is recovered including the (irrelevant)
zero-point energy, which has been omitted in Eq.~\eqref{eq:H}, i.e.,
\begin{equation}
\int d^3 r \, \hat{w}_{\textrm{LA}}({\mathbf{r}},t) =
\sum_{\mathbf{q}}\hbar\omega_{\mathbf{q}} \left( \hat{b}_{\mathbf{q}}^\dag
\hat{b}_{\mathbf{q}}^{} +\frac{1}{2} \right).
\end{equation}
This confirms the consistent definition of acoustic energy density and
acoustic Poynting vector for our model. In the following, when speaking of
the acoustic energy density and the acoustic Poynting vector we will refer to
the expectation value of the corresponding operators, i.e.,
\begin{subequations}
\begin{eqnarray}
w_{\textrm{LA}}({\mathbf{r}},t) &=& \langle \hat{w}_{\textrm{LA}}({\mathbf{r}},t)\rangle ,\\
\mathbf{S}_{\textrm{LA}}({\mathbf{r}},t) &=&
\langle \hat{\mathbf{S}}_{\textrm{LA}}({\mathbf{r}},t)\rangle .
\end{eqnarray}
\end{subequations}
The total energy of the phonon system at time $t$ is obtained from the phonon
Hamiltonian as
\begin{equation}
E_{\textrm{LA}}(t) = \sum_{\mathbf{q}}\hbar\omega_{\mathbf{q}} \langle \hat{b}_{\mathbf{q}}^\dag
\hat{b}_{\mathbf{q}}^{} \rangle .
\label{eq:phonon_energy}
\end{equation}
As soon as the QD is no more in its ground state there is an additional
energy contribution resulting from the exciton-phonon interaction which is
given by
\begin{equation}
E_{\textrm{X-LA}}(t) = \sum_{\mathbf{q}}\hbar\left(g_{\mathbf{q}}
\bigl\langle \hat{b}_{\mathbf{q}} |x\rangle\langle x| \bigr\rangle +
g_{\mathbf{q}}^\ast \bigl\langle \hat{b}_{\mathbf{q}}^\dag |x\rangle\langle x| \bigr\rangle \right).
\label{eq:interaction_energy}
\end{equation}

In general, the optical excitation of the QD leads to the generation of a
coherent amplitude of the lattice displacement described by the expectation
value
\begin{equation}
{\mathbf{u}}({\mathbf{r}},t) = \langle \hat{{\mathbf{u}}}({\mathbf{r}},t)\rangle .
\end{equation}
This coherent phonon amplitude essentially describes the dynamics of a
classical strain field. In special cases, e.g., in the case of excitation by
an ultrafast pulse with a pulse area given by an odd multiple of $\pi$, the
phonons are generated in a purely coherent state,\cite{reiter2011gen} which
means that
\begin{equation}
\langle \hat{b}_{\mathbf{q}}^2 \rangle = \langle \hat{b}_{\mathbf{q}}
\rangle^2\quad \mbox{and} \quad \langle (\hat{b}_{\mathbf{q}}^\dag)^2 \rangle
= \langle \hat{b}_{\mathbf{q}}^\dag \rangle^2 .
\end{equation}
In general, however, i.e., for arbitrary pulse area and pulse duration, the
phonons are not generated in a purely coherent state and both coherent and
incoherent phonons are emitted. Then the acoustic energy density and the
acoustic Poynting vector can be separated into a coherent and an incoherent
part, respectively, according to
\begin{subequations}
\begin{eqnarray}
\left( S_{\textrm{LA}}^{\textrm{coh}} \right)_i &=& -\rho c_l^2 \sum_{j}
\frac{\partial u_j({\mathbf{r}},t)}{\partial t} \frac{\partial u_j({\mathbf{r}},t)}{\partial x_i} ,
\label{eq:poynting-LA-coh}
\\
\mathbf{S}_{\textrm{LA}}^{\textrm{incoh}} &=& \mathbf{S}_{\textrm{LA}} -
\mathbf{S}_{\textrm{LA}}^{\textrm{coh}} ,
\label{eq:poynting-LA-tot}
\end{eqnarray}
\end{subequations}
and correspondingly for the energy density.

Because of the spherical symmetry of the system the acoustic Poynting vector
has only a radial component and it depends only on the distance $r$ from the
QD center, i.e.,
${\mathbf{S}}_{\textrm{LA}}({\mathbf{r}},t)=S_{\textrm{LA}}(r,t)
{\mathbf{e}}_r$ with ${\mathbf{e}}_r$ being the radial unit vector. Due to
energy conservation, it decays $\sim r^{-2}$. To compensate for this
geometrical decay, when discussing its spatio-temporal dynamics we will plot
a scaled acoustic Poynting vector
\begin{equation}
\tilde{S}_{\textrm{LA}}(r,t)=\left(\frac{r}{1\,{\textrm{nm}}}\right)^2
S_{\textrm{LA}}(r,t),
\label{eq:stilde}
\end{equation}
which is defined in such a way that at a radius of 1~nm its value agrees with
the true acoustic Poynting vector. By integrating the acoustic Poynting
vector over the surface of a sphere with radius $r$ as well as over the past
time, we obtain the total elastic energy $\mathcal{E}_{\textrm{LA}}(r,t)$
which has passed through that spherical surface up to time $t$:
\begin{equation}
\mathcal{E}_{\textrm{LA}}(r,t) = 4\pi r^2 \int_{-\infty}^{t}
S_{\textrm{LA}}(r,t^{\prime})\,dt^{\prime}.
\label{eq:Poynting_int}
\end{equation}

For the same symmetry reason as above and because of the restriction to LA
phonons, also the displacement field associated with the coherent phonons has
only a radial component and it only depends on the distance $r$ from the QD
center, i.e., $\langle \hat{{\mathbf{u}}}({\mathbf{r}},t)\rangle = u(r,t)
{\mathbf{e}}_r$. Since the displacement of an emitted spherical phonon wave
packet decays $\sim r^{-1}$, when showing the results in the figures below we
will again compensate for this geometrical decay by plotting the scaled
quantity
\begin{equation}
\tilde{u}(r,t)=\left( \frac{r}{1\,{\textrm{nm}}} \right) u(r,t).
\label{eq:utilde}
\end{equation}
The coherent part of the acoustic energy density and the acoustic Poynting
vector can be directly expressed in terms of the radial displacement
according to
\begin{subequations}
\begin{eqnarray}
w_{\textrm{LA}}^{\textrm{coh}}(r,t)&=& \frac{1}{2} \rho \left[ \left(
\frac{\partial u}{\partial t} \right)^2 + c_l^2 \left(
\frac{\partial u}{\partial r} \right)^2 \right] \\
S_{\textrm{LA}}^{\textrm{coh}}(r,t) &=& -\rho c_l^2 \frac{\partial u}{\partial t}
\frac{\partial u}{\partial r} .
\label{eq:Scoh}
\end{eqnarray}
\end{subequations}

\begin{table}
\caption{\label{tab:table1} Material parameters for GaAs}
\begin{ruledtabular}
\begin{tabular}{l l l}
material density & $\rho$ & 5.37~g/cm$^3$\\
longitudinal sound velocity & $c_l$ & 5110~m/s\\
electron deformation potential & $D_{\textrm{e}}$ & 7.0~eV\\
hole deformation potential & $D_{\textrm{h}}$ & -3.5~eV\\
\end{tabular}
\end{ruledtabular}
\end{table}

While in the case of excitation by ultrafast laser pulses the coupled
QD-phonon dynamics can be calculated analytically,\cite{vagov2002ele} for
pulses with finite duration and for cw excitation no exact analytical results
are known. Therefore we use a numerical calculation on the level of a
fourth-order correlation expansion \cite{rossi2002the,krugel2006bac} which
has been shown to provide very reliable results in the parameter range
studied here.\cite{glassl2011lon} From the calculations we obtain the
coherent phonon mode amplitudes $\langle \hat{b}_{\mathbf{q}}\rangle$ and
$\langle \hat{b}_{\mathbf{q}}^\dag\rangle$ which, according to
Eq.~\eqref{eq:u}, determine the displacement field
$\langle\hat{\mathbf{u}}(\mathbf{r},t)\rangle$ and thus all other coherent
phonon-related quantities introduced above. In addition we get the
expectation values of products of phonon operators $\langle
\hat{b}_{\mathbf{q}}\hat{b}_{{\mathbf{q}^{\prime}}} \rangle$, $\langle
\hat{b}_{\mathbf{q}}^\dag \hat{b}_{{\mathbf{q}^{\prime}}}^{ } \rangle$ and
$\langle \hat{b}_{\mathbf{q}}^\dag \hat{b}_{{\mathbf{q}^{\prime}}}^\dag
\rangle$, from which the total acoustic energy and the total acoustic
Poynting vector can be calculated. Since we are interested in excitation
induced properties of the phonons, we restrict ourselves to the temperature
$T=0$~K. Initially both the exciton and the phonon system are taken to be in
their respective ground state. The material parameters taken in the
calculations are summarized in Table~\ref{tab:table1}.

\section{Pulsed excitation}\label{sec:pulse}

Let us first consider the case of excitation by a Gaussian laser pulse with
pulse area $A$ and pulse width $\tau$, where the Rabi frequency entering in
Eq.~(\ref{eq:rabi}) is given by
\begin{equation}
\Omega(t) =  \frac{A}{\tau\sqrt{2\pi}}\exp \left(-\frac{t^2}{2\tau^2} \right).
\label{eq:pulse}
\end{equation}
In an ideal two-level system without coupling to phonons, the excitation by a
pulse with pulse area $A=\pi$ completely inverts the system. In the present
case, due to the exciton-phonon coupling, the excitation of the electronic
system is in general associated with an excitation of the phonon system
resulting in a dephasing of the electronic system and the generation of
coherent and incoherent phonons.\cite{krugel2006bac} Both the dephasing and
the phonon generation have been found to strongly depend on the pulse width
$\tau$.\cite{krugel2005the,krugel2006bac,machnikowski2007qua} In the
following we will review the acoustic phonon dynamics and in particular
analyze the energy transfer to and energy transport by the acoustic phonons
in the case of excitation by pulses with different pulse areas and pulse
widths.

\begin{figure}
\includegraphics[width=\columnwidth]{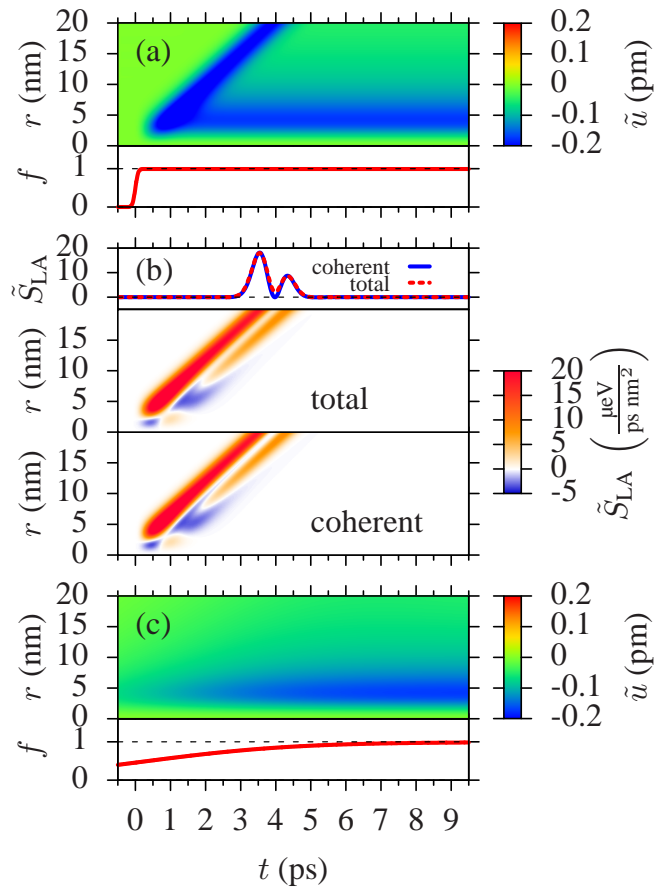}
\caption{(Color online) (a) Coherent lattice displacement $\tilde{u}(r,t)$
[scaled according to Eq.~\eqref{eq:utilde}] as a function of time $t$
and distance from the QD center $r$ (upper panel) and occupation $f$ of the
exciton state as a function of time (lower panel) for the case of excitation by
a laser pulse with pulse area $A=\pi$ and pulse width $\tau=100\,{\textrm{fs}}$. (b)
Total and coherent acoustic Poynting vector of the LA phonons $\tilde{S}_{\textrm{LA}}(r,t)$
[scaled according to Eq.~\eqref{eq:stilde}] as functions of time $t$
and distance from the QD center $r$ (lower and central panel) and as a function of time at
$r=20\,{\textrm{nm}}$ (upper panel). (c) Same as part (a) but for a $\pi$-pulse with pulse width
$\tau=5\,{\textrm{ps}}$.}
\label{fig:1pi}
\end{figure}

Figure~\ref{fig:1pi} shows for the case of excitation by a $\pi$-pulse with
width $\tau=100$~fs [Fig.~\ref{fig:1pi}(a)] and $\tau=5$~ps
[Fig.~\ref{fig:1pi}(c)] the occupation of the exciton state $f = \bigl\langle
|x\rangle\langle x| \bigr\rangle$ as a function of time (lower panels) and
the coherent lattice displacement $\tilde{u}(r,t)$  as a function of time and
distance from the QD (upper panels). (Note that the plotted displacement has
been scaled according to Eq.~\eqref{eq:utilde}.) While the exciton occupation
in both cases essentially reaches one, the resulting phonon dynamics is very
different. In both cases a negative lattice displacement builds up in the
region of the QD, i.e., the lattice contracts radially towards the QD
(horizontal blue lines at $r \lesssim 5$~nm). In addition, in
Fig.~\ref{fig:1pi}(a) a pronounced negative displacement outside the QD
region shows up (diagonal blue line) while in Fig.~\ref{fig:1pi}(c)
essentially no displacement outside the QD region is seen.

The lattice displacement in the QD region results from the fact that,
according to the linear exciton-phonon coupling in Eq.~\eqref{eq:H}, the
equilibrium positions of the lattice ions are shifted if the electronic
system is in its excited state. The horizontal lines therefore depict the
build-up of the acoustic polaron in the QD region.

As has been discussed in previous studies,\cite{machnikowski2007qua} the two
pulse durations shown in Fig.~\ref{fig:1pi} correspond to two limiting cases:
In the case of the long pulse with $\tau=5$~ps [Fig.~\ref{fig:1pi}(c)], the
temporal evolution occurs on a time scale slower than the characteristic time
scale of the phonons, which can be estimated as the oscillation period of the
most strongly coupled phonon mode, $T_{\textrm{ph}} \approx 2$~ps. In this
case the lattice ions move adiabatically into their new equilibrium
positions. In contrast, in the case of the short pulse
[Fig.~\ref{fig:1pi}(a)] the excitation of the electronic system occurs much
faster than $T_{\textrm{ph}}$, such that the phonons cannot follow
immediately. Instead, the polaron builds up after the pulse and the excess
energy which is released by the polaron formation is emitted in the form of a
phonon wave packet which leaves the QD region with the sound velocity $c_l$
[diagonal blue line in Fig.~\ref{fig:1pi}(a)]. The emitted wave packet has a
temporal width of about 1~ps which is directly related to the bandwidth of
the phonon spectral density.

In Fig.~\ref{fig:1pi}(b) we have plotted the acoustic Poynting vector of the
LA phonons $\tilde{S}_{\textrm{LA}}$ [scaled according to
Eq.~\eqref{eq:stilde}] for the case of excitation by the 100~fs pulse. The
lower and central parts show the coherent and total acoustic Poynting vector,
respectively, as functions of time and distance from the QD, the upper part
shows both quantities as functions of time at the distance of
$r=20\,{\textrm{nm}}$. We observe that the coherent and the total acoustic
Poynting vector exhibit an almost perfect agreement indicating that in this
case the phonons are generated in an almost perfect coherent state. Since the
Poynting vector associated with coherent phonons is determined by the
temporal and spatial derivatives of the displacement [see
Eq.~\eqref{eq:poynting-LA-coh}], the emitted phonon wave packet gives rise to
a double-peak structure with maxima at the rising and falling edge of the
wave packet and a minimum in the center. Only at this minimum a slight
difference between coherent and total Poynting vector is seen. In the case of
excitation by the 5~ps pulse no phonon wave packet is emitted and thus the
acoustic Poynting vector is very small and therefore it is not shown.

The spatio-temporal evolution of the lattice displacement in the case of the
short pulse excitation agrees well with calculations in the $\delta$-pulse
limit discussed in Ref.~\onlinecite{wigger2013flu}, where the exciton and
phonon dynamics can be calculated analytically without any
approximation.\cite{vagov2002ele} In particular, it can be shown that the
excitation by an ultrafast $\pi$-pulse indeed results in the generation of
purely coherent phonons.\cite{sauer2010lat} From the excellent agreement we
can conclude that, on the one hand, the $\delta$-pulse limit is in this
system a reasonable approximation for laser pulses with durations in the
range of hundred femtoseconds, and that, on the other hand, the fourth-order
correlation expansion is an adequate approximation to describe the phonon
dynamics with high precision.

\begin{figure}
\includegraphics[width=\columnwidth]{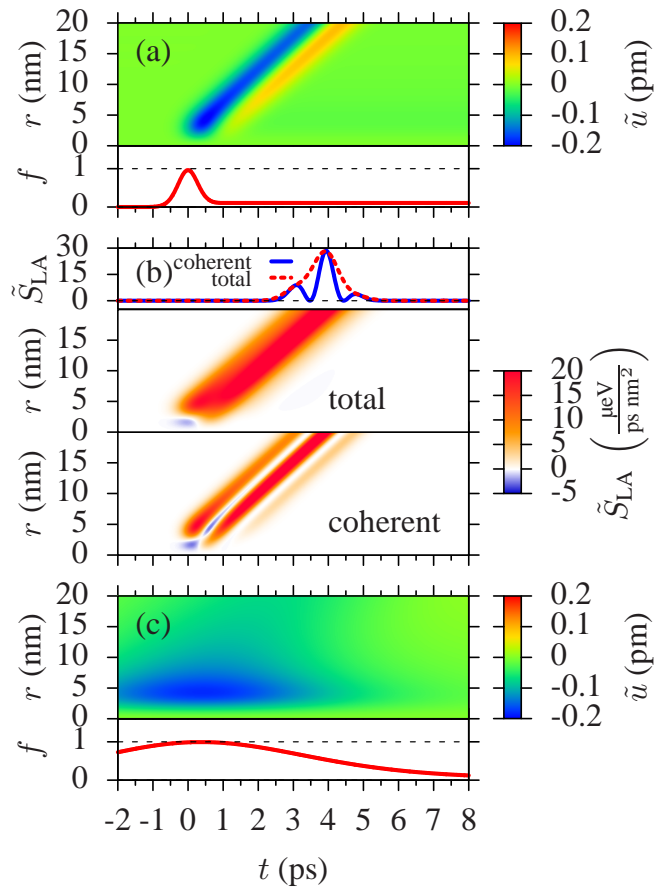}
\caption{(Color online) Same as Fig.~\ref{fig:1pi} but for a pulse area of
$A=2\pi$ and pulse widths $\tau=0.5\,{\textrm{ps}}$ [parts (a) and (b)]
and $\tau=5\,{\textrm{ps}}$ [part (c)].}
\label{fig:2pi}
\end{figure}

When the pulse area of the exciting laser pulse is increased to $A=2\pi$, the
electronic system is first excited and subsequently de-excited again. In an
ideal two-level system the exciton occupation thus returns to zero after the
pulse. Figure~\ref{fig:2pi} shows the exciton and phonon dynamics of the
coupled exciton-phonon system for the case of excitation by a $2\pi$-pulse
with pulse widths (a) $\tau=0.5$~ps and (c) $\tau=5$~ps.

In the case of the long pulse [Fig.~\ref{fig:2pi}(c)] we observe again the
quasi-adiabatic evolution of the exciton-phonon system. With increasing
occupation of the exciton the polaron builds up and with the subsequent
decrease in the occupation also the polaron decays such that after the pulse
no excitation is left, neither in the electronic nor in the phonon system.
The other limiting case of excitation by an ultrafast pulse with $\tau
\lesssim 100$~fs is not shown here, because in this case the electronic
system is so rapidly excited and de-excited that the phonon system cannot
follow and essentially remains in its ground state. For an intermediate pulse
width of $\tau=0.5$~ps [Fig.~\ref{fig:2pi}(a)] we find that the rapidly
rising exciton occupation is associated with the build-up of a polaron and
the emission of a phonon wave packet with a negative displacement (blue
diagonal line). The subsequent decrease of the occupation leads to a decay of
the polaron which, because of the rather short time scale, is again
associated with the emission of a phonon wave packet, now however with a
positive value of the displacement (yellow-red diagonal line). The
irreversibility introduced by the emission of the phonon wave packet gives
rise to the dephasing of the electronic degrees of freedom. As a result, the
exciton occupation does not completely return to zero.

When looking again at the acoustic Poynting vector [Fig.~\ref{fig:2pi}(b)],
we now observe a pronounced difference between the total energy flow and the
energy flow associated with coherent phonons. The coherent acoustic Poynting
vector exhibits three peaks located at the positions of maximal slopes of the
displacement profile. The total Poynting vector instead is a smooth function
of time which forms an envelope over the coherent part showing that between
the peaks resulting from the coherent phonons there is still an energy
transport but now related to incoherent phonons.

\begin{figure}
\includegraphics[width=\columnwidth]{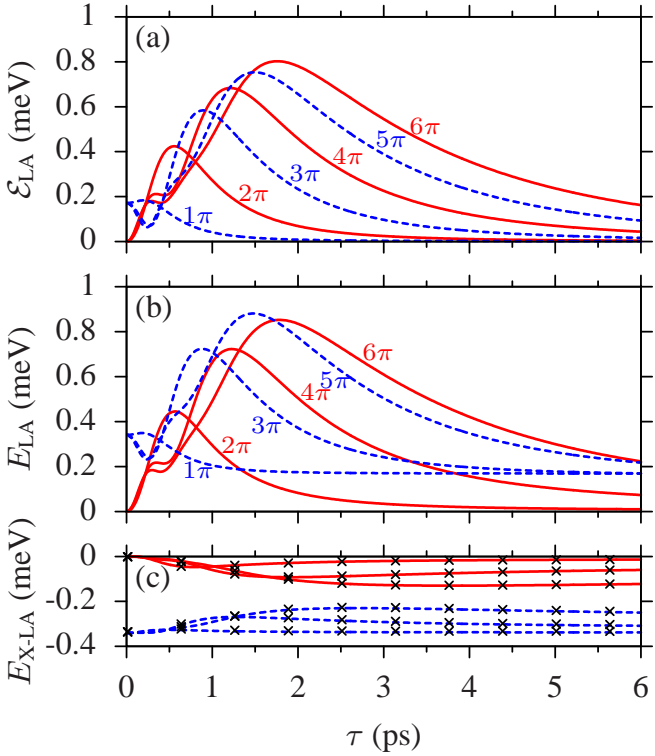}
\caption{(Color online) (a) Total energy carried away by the phonons, (b)
total energy transferred to the phonon system, and (c) final exciton-phonon
interaction energy as a
function of the pulse width $\tau$ of the exciting laser pulse
for pulses with pulse areas $A=n\pi$ for $n=1,\dots,6$. For the blue dashed curves
in (c) (pulse areas of odd multiples of $\pi$) the pulse
area increases from bottom to top; for the red solid lines (pulse areas of
even multiples of $\pi$) the pulse area increases from top to bottom. The symbols
in (c) show the energy $-2f \hbar\omega_{\textrm{pol}}$ with the calculated exciton
occupation $f$ at the respective pulse width and pulse area.}
\label{fig:E_tau}
\end{figure}

The energy dissipation to the phonon system after excitation by pulses with
different pulse areas is summarized in Fig.~\ref{fig:E_tau}(a), where we have
plotted the total elastic energy
${\mathcal{E}}_{\textrm{LA}}(r=20\,\textrm{nm},t=\infty)$, which flows
through the surface of a sphere with 20~nm radius after excitation by a pulse
with pulse area $A=n\pi$ as a function of the pulse width $\tau$ for
$n=1,\dots,6$. The radius $r=20$~nm has been chosen such that the polaron
contribution to the energy flow is negligible. Outside of the polaron region
the total energy ${\mathcal{E}}_{\textrm{LA}}(r,t=\infty)$ flown through the
surface of a sphere with radius $r$ is in fact independent of $r$.

Let us first concentrate on the two curves corresponding to pulse areas $\pi$
and $2\pi$. While in the limit of long pulses for both pulse areas the
adiabatic limit is reached, in which the phonon dynamics is restricted to the
QD region and no energy is emitted, for short pulses there are pronounced
differences. In the case of excitation by a $\pi$-pulse the emitted energy
becomes maximal in the limit of ultrafast pulses and decays monotonically
with increasing pulse width. In contrast, for a $2\pi$-pulse the dynamics of
the exciton is decoupled from the phonon dynamics in the limit of ultrafast
pulses. Hence, there is no energy emitted. Here, the emitted acoustic energy
becomes maximal for pulse widths of about $0.5$~ps.

\begin{figure*}
\includegraphics[width=1.5\columnwidth]{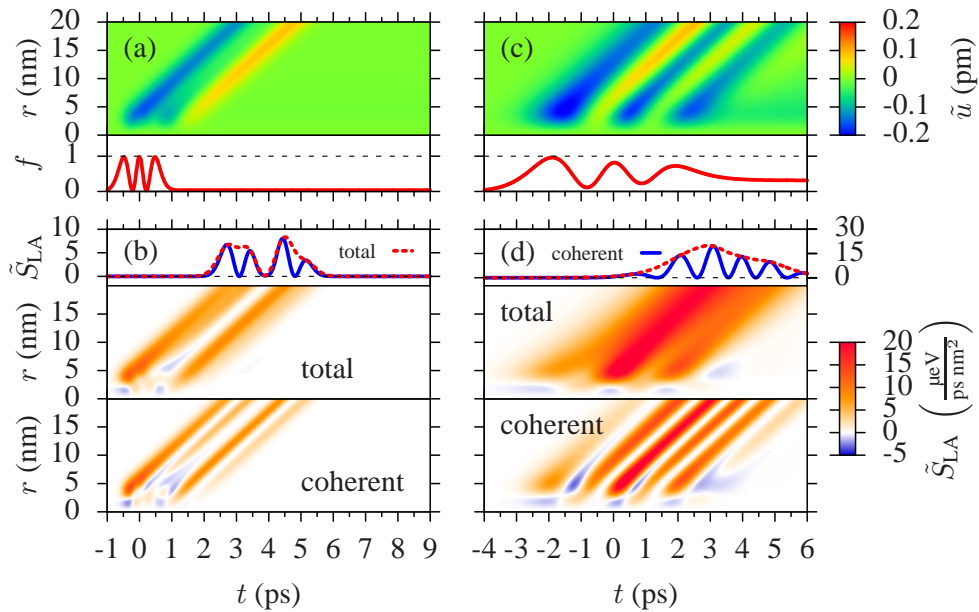}
\caption{(Color online) Same as Fig.~\ref{fig:1pi} but for a pulse area of
$A=6\pi$ and pulse widths $\tau=0.5\,{\textrm{ps}}$ [parts (a) and (b)] and
$\tau=2\,{\textrm{ps}}$ [parts (c) and (d)].} \label{fig:6pi}
\end{figure*}

When increasing the pulse area $A$ beyond $2\pi$, the electronic system
performs multiple Rabi rotations. The characteristic time scale of the
electronic evolution increasingly deviates from the pulse width $\tau$.
Instead it is determined by the instantaneous Rabi period $2\pi/\Omega(t)$
which then determines the coupling to the phonons. In Fig.~\ref{fig:6pi} the
exciton and phonon dynamics are shown for the case of excitation by pulses
with a pulse area of $A=6\pi$ and a pulse width of $\tau=0.5$~ps
[Figs.~\ref{fig:6pi}(a) and (b)] and $\tau=2$~ps [Figs.~\ref{fig:6pi}(c) and
(d)]. The exciton occupation performs three Rabi flops and, in the case of
the short pulse, essentially returns to zero while for the longer pulse an
occupation of about 0.3 remains which already indicates a rather strong
dephasing. In the case of the short pulse [Fig.~\ref{fig:6pi}(a)] we observe
the emission of a wave packet with negative amplitude resulting from the
initial increase of the occupation and a wave packet with positive amplitude
caused by the de-excitation during the trailing edge of the pulse. The Rabi
oscillations in between occur on a shorter time scale of the order of 100~fs,
which is faster than the typical phonon time scale, and therefore they are
almost decoupled from the phonons. In contrast, for the longer pulse
[Fig.~\ref{fig:6pi}(c)] the period of the Rabi oscillations is in the range
where the exciton-phonon coupling is rather efficient. We observe the
emission of three pronounced wave packets with a negative amplitude
associated with a rising exciton occupation and in between two wave packets
with positive amplitude associated with the decreasing occupation. The final
decay is rather weak due to the strong dephasing and therefore does not
anymore lead to a clear coherent phonon emission.

Figures~\ref{fig:6pi}(b) and (d) illustrate the energy flow associated with
the generated phonons for the two pulse durations. In the case of the longer
pulse, the total acoustic Poynting vector again forms an envelope over the
peaks associated with the emission of coherent phonons. For the shorter
pulse, on the other hand, the total energy flow goes to zero at the center of
the wave packet. Here the decoupling between exciton and phonons due to the
fast exciton dynamics also inhibits the generation of incoherent phonons. The
total acoustic Poynting vector now exhibits a double peak structure.

Let us now come back to the discussion of the emitted energy. In addition to
the cases of $A=\pi$ and $A=2\pi$, Fig.~\ref{fig:E_tau}(a) shows the total
energy transported away from the QD by the phonon wave packets also for
pulses with higher pulse areas. For even multiples of $\pi$ all curves start
at zero while for odd multiples all curves start at the same finite value.
This is due to the fact that for very short pulses the exciton dynamics is
decoupled from the phonons and the subsequent phonon dynamics only depends on
the final exciton occupation, which is zero for even multiples and one for
odd multiples of $\pi$.  The maxima of the curves increase with increasing
pulse area and they shift to higher values of the pulse width. This shift
again reflects the fact that the efficiency of the coupling is determined by
the Rabi frequency instead of the pulse width. As can be seen from
Eq.~\eqref{eq:pulse}, a larger pulse area $A$ requires a larger pulse width
$\tau$ to reach the same peak Rabi frequency $\Omega_{\textrm{max}} = (A/\tau
\sqrt{2\pi})$. The increase in the emitted energy at the maximum of the curve
with increasing pulse area is due to the fact that each Rabi flop gives rise
to the emission of a pair of wave packets. Since, however, as seen in
Fig.~\ref{fig:6pi}(b), the amplitude of subsequent Rabi flops decreases due
to an increasing dephasing, the maximally emitted energy does not grow
linearly with the pulse area but in a sub-linear way.

Interestingly, all curves for $A \ge 3\pi$ exhibit a local minimum at short
pulse widths before reaching their maximal value. This minimum results from
the fact that at these short pulses there is a significant destructive
interference between wave packets with negative amplitude emitted during the
rise of the excitonic occupation and those with positive amplitude emitted
during the decrease of the occupation. This destructive interference reduces
the energy transported away from the QD by the phonon wave packets. In
particular in the case of pulses with pulse areas given by odd multiples of
$\pi$ this leads to a rapid initial decrease of the emitted phonon energy for
very short pulses.

Figure~\ref{fig:E_tau}(b) shows the total energy of the phonon system
$E_{\textrm{LA}}(t=\infty)$ calculated according to
Eq.~\eqref{eq:phonon_energy}. For pulse areas given by even multiples of
$\pi$ the curves in Fig.~\ref{fig:E_tau}(b) are very similar to the ones in
Fig.~\ref{fig:E_tau}(a). In these cases there is only a small exciton
occupation after the pulse and thus there is only a small polaron
contribution remaining. Therefore, almost all of the energy transferred to
the phonons is transported away from the QD region. In the case of pulse
areas given by odd multiples of $\pi$ the QD essentially remains in the
excited state after the pulse which is associated with a polaronic lattice
distortion in the QD region. The total energy of the phonons is therefore
larger than the energy transported away, the difference being given by the
energy of the polaron.

This interpretation is confirmed and quantified by looking at the interaction
energy calculated according to Eq.~\eqref{eq:interaction_energy}, which is
plotted in Fig.~\ref{fig:E_tau}(c). In the ultrafast limit the interaction
energy is $-2\hbar\omega_{\textrm{pol}}$ for all odd multiples of $\pi$ while
it vanishes for all even multiples. This is because in the former case the
exciton occupation after the pulse is given by $f=1$ while in the latter case
it is $f=0$. For larger pulses, due to the dephasing during the pulse, there
is a non-zero final occupation after pulses with an integer multiple of $\pi$
and a non-unit occupation after pulses with an odd multiple of $\pi$. Thus
the interaction energy is in general between the limiting cases found for
ultrafast pulses. For long pulses we approach the adiabatic regime where
again these values are reached. This is seen in particular in the cases of
$\pi$ and $2\pi$ pulses while for pulses with higher pulse areas much longer
pulse widths are required to reach this regime. We find that the interaction
energy in all cases is given by $-2f \hbar\omega_{\textrm{pol}}$ [symbols in
Fig.~\ref{fig:E_tau}(c)]. Furthermore, the energies shown in
Fig.~\ref{fig:E_tau} satisfy the relation
\begin{equation}
E_{\textrm{LA}}-{\mathcal{E}}_{\textrm{LA}}=-\frac{1}{2} E_{\textrm{X-LA}},
\end{equation}
confirming again that indeed except for the energy residing in the polaron
all the remaining energy is emitted from the QD.

From the results discussed so far we can conclude that the instantaneous Rabi
frequency has a significant impact on the phonon emission of the QD. On the
one hand, very slow excitations adiabatically build up and remove the polaron
and, on the other hand, very fast Rabi oscillations lead to a decoupling of
exciton and phonon dynamics. Only for intermediate Rabi frequencies we find
an efficient carrier-phonon coupling leading to a pronounced transfer of
energy to the phonons and a subsequent transport of energy away from the QD
region. To analyze this resonance phenomenon in more detail, in the next
section we will study the case of excitations in which the Rabi frequency is
fixed after the optical field has been switched on.

\section{Continuous excitation switched on instantaneously}\label{sec:cw}

\begin{figure*}[t]
\includegraphics[width=1.8\columnwidth]{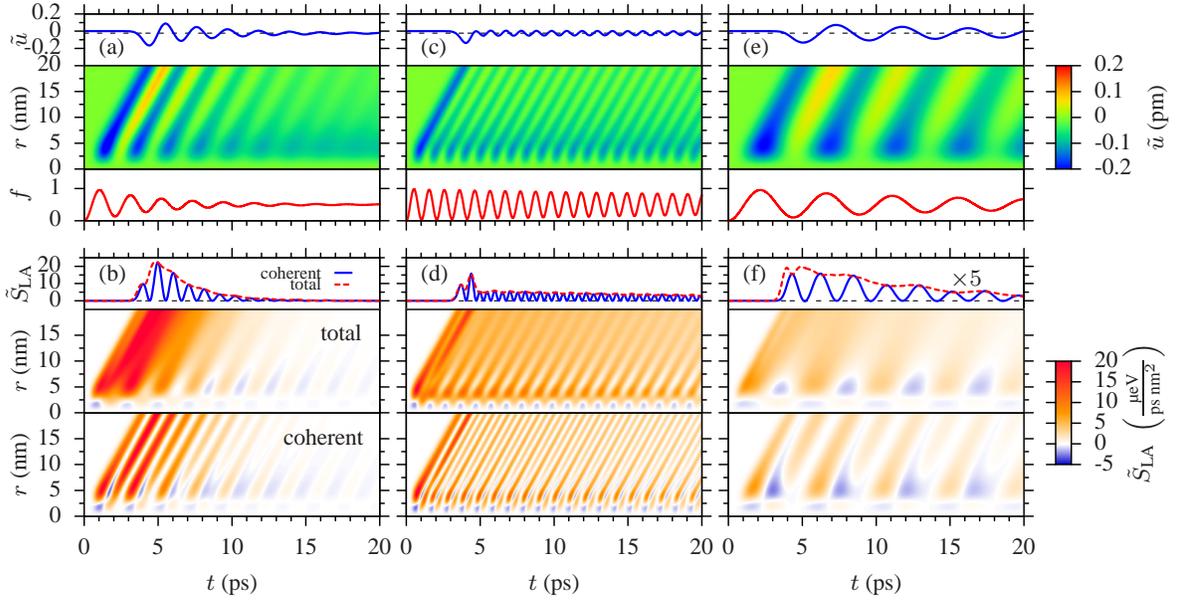}
\caption{(Color online) (a),(c),(e) Coherent lattice displacement
$\tilde{u}(r,t)$ [scaled according to Eq.~\eqref{eq:utilde}] as a function of
time $t$ and distance $r$ from the QD center $r$ (central panels) and as a
function of time at $r=20\,{\textrm{nm}}$ (upper panels) and occupation $f$
of the exciton state as a function of time (lower panels) for a cw excitation
switched on at time $t=0$. (b),(d),(f) Total and coherent acoustic Poynting
vector of the LA phonons $\tilde{S}_{\textrm{LA}}(r,t)$ [scaled according to
Eq.~\eqref{eq:stilde}] as functions of time $t$ and distance from the QD
center $r$ (central and lower panels) and as a function of time at
$r=20\,{\textrm{nm}}$ (upper panels). The Rabi frequencies are (a),(b)
$\Omega_{\textrm{R}}=3\,{\textrm{ps}}^{-1}$; (c),(d)
$\Omega_{\textrm{R}}=6\,{\textrm{ps}}^{-1}$; (e),(f)
$\Omega_{\textrm{R}}=1.5\,{\textrm{ps}}^{-1}$. } \label{fig:cw}
\end{figure*}

In this section we consider a light field that is switched on instantaneously
at $t=0$. After the switching-on the amplitude of the light field remains
constant, so that the Rabi frequency in Eq.~(\ref{eq:rabi}) is
time-independent, i.e., $\Omega(t) = \Omega_{\textrm{R}}$. The associated
Rabi period is $T_{\textrm{R}}=2\pi /\Omega_{\textrm{R}}$. A continuously
driven two-level system which is coupled to phonons performs damped Rabi
oscillations. In the limit of long times the system reaches a stationary
state, where in the case of resonant driving the occupation of the exciton is
$f=0.5$ and the polarization depends on the coupling to the
phonons.\cite{glassl2011lon} Here we again want to focus on the properties of
the generated phonons.

First we analyze the case when the Rabi frequency is equal to the frequency
where the phonon spectral density has its maximum, i.e.,
$\Omega_{\textrm{R}}=\omega_{\textrm{ph}}=3\,{\textrm{ps}}^{-1}$.
Figure~\ref{fig:cw}(a) shows in the lower panel the occupation of the exciton
state $f$ as a function of time after switching on the light field. One
clearly observes Rabi oscillations with a period of
$T_{\textrm{R}}\approx2\,{\textrm{ps}}$ that are strongly damped in time.
Already after $20\,{\textrm{ps}}$ a stationary occupation of $f=0.5$ is
reached. In the lattice displacement $\tilde{u}$ shown in the central panel a
series of diagonal lines with alternating sign appears reflecting emitted
phonon wave packets. The amplitudes are getting smaller with every line. For
long times the lines almost vanish and only a weak horizontal line around $r
\approx 5\,{\textrm{nm}}$ remains. This line represents the polaron which is
reduced compared to Fig.~\ref{fig:1pi} due to the reduced exciton occupation
of $f=0.5$ in the stationary state. For every Rabi flop a pair of wave
packets, one with negative and one with positive amplitude, is emitted. While
the first wave packets have large amplitudes, the amplitudes get smaller
quite fast. All the features of the wave packets can also be seen in the
upper panel where we show $\tilde{u}$ as a function of time for the fixed
distance of $r=20\,{\textrm{nm}}$. We notice that the first wave packet
arrives at $t=4\,{\textrm{ps}}$, which is the time the phonons need to cover
a distance of $20\,{\textrm{nm}}$. Both, in the oscillation period and the
damping behavior we see a strong analogy between the lattice displacement and
the Rabi oscillation, reflecting the tight connection between the two
systems. The small lattice displacement remaining for long times is the tail
resulting from the polaronic displacement, which is still slightly visible
even at a distance of $r=20$~nm. It agrees with the value of the polaronic
displacement at this distance calculated analytically for the case of
ultrafast excitation \cite{wigger2013flu} included as dashed line in the
plot, which demonstrates that the lattice distortion associated with the
polaron is an intrinsic feature of the system and independent of the details
of excitation.

Figure~\ref{fig:cw}(b) shows the acoustic Poynting vector
$\tilde{S}_{\textrm{LA}}$ scaled according to Eq.~\eqref{eq:stilde}. The
lower panel displays the coherent part and the central panel the total value,
both plotted as functions of time and distance from the QD. The upper panel
shows the coherent as well as the total value at a distance of 20~nm from the
QD as a function of time. Like in the case of pulsed excitation we find peaks
in the coherent part of the energy at the edges of the coherent phonon wave
packets while the total acoustic Poynting vector is a smooth function of time
and forms an envelope over the coherent part. Both the coherent and
incoherent phonon emission are strongly damped on the same time scale as the
oscillations in the exciton occupation. While in the region far from the QD
the energy flow is always positive, i.e., directed away from the QD, close to
the QD at distances $r\lesssim 5\,{\textrm{nm}}$ an oscillation of the
acoustic Poynting vector between negative and positive values appears. This
pattern shows the effect of the creation and destruction process of the
polaron in the QD area. The steady contraction and relaxation of the lattice
atoms leads to an oscillatory behavior of the energy in an area close to the
QD which is obviously not only associated with the oscillatory energy
transfer between exciton and phonons but also with a spatial flow of energy
in the phonon system to and from the QD region.

Let us now consider the case of a Rabi frequency twice as large as previously
studied, i.e., $\Omega_{\textrm{R}}=6\,{\textrm{ps}}^{-1}$.
Figure~\ref{fig:cw}(c) shows in the lower panel the corresponding Rabi
oscillation, which has an oscillation period of $1\,{\textrm{ps}}$. We can
see that now the oscillation is only slightly damped. The lattice
displacement in the central and upper panels shows a sequence of phonon wave
packets emitted with the same frequency as the Rabi oscillations. Compared to
the first wave packet the remaining wave packets of this sequence have only
small amplitudes. The first wave packet is emitted at $t=0$ and has a width
of approximately $1\,{\textrm{ps}}$. It is a result of the instantaneous
switching-on of the light field. During the rest of the time only weak wave
packets are emitted. This is a result of the high driving frequency of the
exciton system, which is too fast for the lattice atoms to follow.

Looking now at the acoustic Poynting vector in Fig.~\ref{fig:cw}(d), a
constant series of diagonal lines with significant amplitudes is visible both
in the coherent part and the total energy flow. Although the emitted wave
packets are quite weak, they carry a remarkable amount of energy with them.
In addition, the amplitudes are almost undamped on the time scale shown here.
In the region close to the QD we again see the oscillatory behavior of the
energy flux.

Finally we take $\Omega_{\textrm{R}}=1.5\,{\textrm{ps}}^{-1}$, i.e., half the
value of Figs.~\ref{fig:cw}(a),(b). The results are shown in the right column
of Fig.~\ref{fig:cw}. The Rabi oscillation in the occupation plotted in the
lower panel of Fig.~\ref{fig:cw}(e) has an oscillation period time of about
$4\,{\textrm{ps}}$ and only a moderate damping. The lattice displacement in
the central and upper panel shows again a sequence of wave packets emitted
from the dot. The wave packets are clearly visible over the whole time window
and have the same damping as the Rabi oscillation. The acoustic Poynting
vector in Fig.~\ref{fig:cw}(f) has the same features as in the two cases
discussed before, but in comparison to the amplitude of the lattice
displacement in Fig.~\ref{fig:cw}(a) it is rather weak. This shows that,
because of the much lower temporal derivative of the displacement, these
broad wave packets carry only a small amount of energy.

\begin{figure}
\includegraphics[width=\columnwidth]{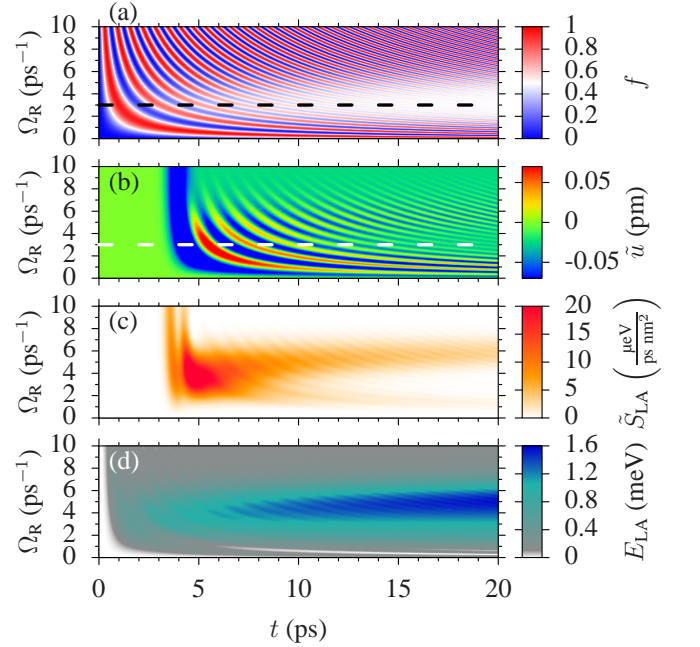}
\caption{(Color online) Exciton and phonon dynamics as a function of time $t$
and Rabi frequency $\Omega_{\textrm{R}}$ for a cw excitation switched on at time $t=0$.
(a) Exciton occupation $f$,
(b) scaled lattice displacement $\tilde{u}$ at a distance of $r=20\,{\textrm{nm}}$, (c)
scaled total acoustic Poynting vector $\tilde{S}_{\textrm{LA}}$ at $r=20\,{\textrm{nm}}$, and (d)
total energy of the phonon system $E_{\textrm{LA}}$.}
\label{fig:cw_omega}
\end{figure}

After discussing these three examples of excitation with a Rabi frequency at,
above, and below the phonon resonance frequency, let us now summarize these
findings by discussing the dependence of the phonon emission properties on
the Rabi frequency. For this purpose, in Fig.~\ref{fig:cw_omega}
characteristic variables of the exciton and phonon system are plotted as
functions of time $t$ and Rabi frequency $\Omega_{\textrm{R}}$.
Figure~\ref{fig:cw_omega}~(a) shows the exciton occupation $f$. As expected,
the electronic system performs damped oscillations between the ground state
and the exciton state with Rabi frequency $\Omega_{\textrm{R}}$, the damping
being strongly dependent on the Rabi frequency. The dashed line marks the
resonance frequency $\Omega_{\textrm{R}}=3\,{\textrm{ps}}^{-1}$, where the
spectral density has its maximum. Indeed, at this Rabi frequency the
dephasing is most efficient. In a region around this resonant Rabi frequency
the oscillations vanish very fast. For long times the occupation $f$ reaches
a stationary value of $0.5$. For frequencies outside this region the Rabi
oscillations are rather stable on the time scale considered here.

To analyze the properties of the emitted phonon wave packets we focus on a
fixed distance of $r=20\,{\textrm{nm}}$. At this distance the influence of
the polaron is negligible, so that the effects of the emitted wave packets
can clearly be seen. Figure~\ref{fig:cw_omega}(b) shows the lattice
displacement, where the dashed line again marks
$\Omega_{\textrm{R}}=3\,{\textrm{ps}}^{-1}$. The first wave packet shows up
at $t\approx 4\,{\textrm{ps}}$, which is attributed to the time that the
phonons need to cover a distance of $20\,{\textrm{nm}}$ after they have been
created at $t=0$. This wave packet is very pronounced and appears as soon as
the Rabi frequency is sufficiently high. Furthermore, the shape of this wave
packet is almost independent of $\Omega_{\textrm{R}}$. As seen in
Fig.~\ref{fig:cw}, this wave packet is a fingerprint of the instantaneous
switching-on of the light field and therefore not sensitive to the Rabi
frequency itself.

In agreement with the behavior of the Rabi oscillations, the amplitudes of
the emitted phonon wave packets are strongly damped around the resonance
frequency $\Omega_{\textrm{R}}=3\,{\textrm{ps}}^{-1}$ showing a coincidence
in the damping behavior between carriers and phonons. For frequencies smaller
than $3\,{\textrm{ps}}^{-1}$ the exciton creation and destruction is slower
than the phonon dynamics, so that the emission of broad wave packets is
possible [cf. Fig.~\ref{fig:cw}(f)]. An efficient phonon emission is also
possible when the Rabi frequency exceeds the resonance frequency, but in this
case the repetition rate of the wave packets is much higher than in the low
frequency limit. If the Rabi frequency is increased even more, i.e., for
$\Omega_{\textrm{R}}\gtrsim 8\,{\textrm{ps}}^{-1}$, the phonons become
essentially decoupled from the carrier dynamics. This is confirmed by the
vanishing amplitudes of the lattice displacement after the pronounced wave
packet emitted at $t=0$. For the exciton occupation this decoupling leads to
almost undamped Rabi oscillations, leading to the reappearance of Rabi
rotations when plotted as a function of the pulse area.\cite{vagov2007non}

To analyze the energy transport associated with the generated phonons, in
Fig.~\ref{fig:cw_omega}(c) the scaled total acoustic Poynting vector
$\tilde{S}_{\textrm{LA}}$ at the distance $r=20\,{\textrm{nm}}$ is shown. The
acoustic Poynting vector also shows a pronounced damping behavior in the
resonance region around $3\,{\textrm{ps}}^{-1}$, where the energy flow decays
within a few picoseconds after the first wave packet arrives. In contrast,
the energy emission lasts much longer if the Rabi frequency is slightly
increased above the phonon resonance. Especially for Rabi frequencies around
$\Omega_{\textrm{R}}=5\,{\textrm{ps}}^{-1}$ we observe a significant and
stable energy flux for the times shown here. Moving to smaller values of
$\Omega_{\textrm{R}}\approx1\,{\textrm{ps}}^{-1}$, the energy flux also lasts
quite long. But despite the pronounced lattice displacements occurring at
these Rabi frequencies, the wave packets carry only little energy because of
the small velocities of the displacement associated with the low frequencies.
This can be seen in the comparably small values of the acoustic Poynting
vector.

Another illustrative quantity to characterize the phonon emission efficiency
is the total energy $E_{\textrm{LA}}(t)$ transferred to the phonons up to the
time $t$, which is shown in Fig.~\ref{fig:cw_omega}(d). While the phonon
energy reaches its final value within the shortest time at the resonance
frequency $\Omega_{\textrm{R}}=3\,{\textrm{ps}}^{-1}$, the overall maximal
energy emission within the time span of $20$~ps shown here is found around
$\Omega_{\textrm{R}}=5\,{\textrm{ps}}^{-1}$. For resonant coupling the energy
transfer is strongest for short times, but vanishes quickly as the exciton
system reaches its stationary state. Instead, for higher Rabi frequencies the
phonon emission is present for a longer time. Although each wave packet
carries a smaller amount of energy than at resonance, the total energy adds
up to a larger value. At low Rabi frequencies $\Omega_{\textrm{R}}\lesssim
1\,{\textrm{ps}}^{-1}$ we observe a bright line starting at about $6$~ps and
a second one starting at about 16~ps indicating that there is essentially no
energy transfer at these combinations of time and Rabi frequency. At these
low Rabi frequencies the system is close to the adiabatic regime where the
polaron is reversibly created and destroyed. Indeed, the bright lines
correspond to completed $2\pi$ and $4\pi$-rotations and thus to a vanishing
exciton occupation, as can be seen by comparison with
Fig.~\ref{fig:cw_omega}(a).

\begin{figure}
\includegraphics[width=\columnwidth]{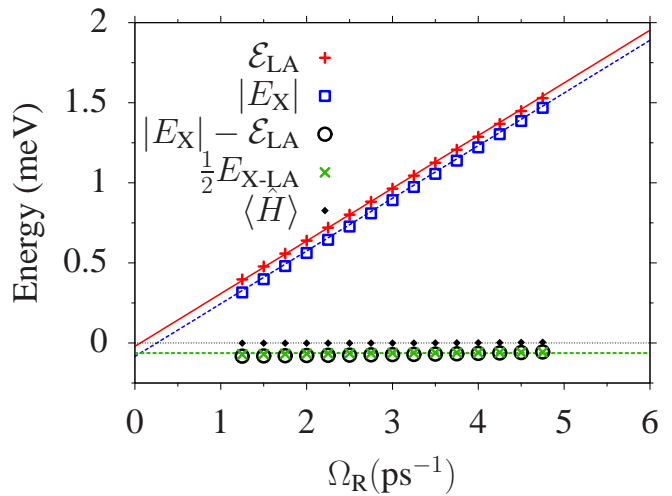}
\caption{(Color online) Symbols show different energy contributions as labeled in the Figure
as a function of the Rabi frequency $\Omega_{\textrm{R}}$ for the case of a cw excitation
switched on at $t=0$. All quantities are calculated at the time
$t=400\,{\textrm{ps}}$, i.e., when the QD has reached the steady state. The
lines are explained in the text.}
\label{fig:cw_energy}
\end{figure}

As is evident from Fig.~\ref{fig:cw_omega}(a), in a region of Rabi
frequencies around the resonant value the exciton occupation reaches its
final value of $0.5$ within the time window of 20~ps shown in the Figure.
Then the whole system, although continuously driven, reaches a steady state
and the energy transferred to the phonons saturates. With increasing distance
from the resonance condition the steady state is reached at increasingly
longer times.

Let us finally investigate quantitatively the energy transfer to the phonon
system after the steady state has been reached. For this purpose we have
extended the calculations to a time window of 400~ps. During this time for
all Rabi frequencies between $\Omega_{\textrm{R}} \approx
1\,{\textrm{ps}}^{-1}$ and $\Omega_{\textrm{R}} \approx
5\,{\textrm{ps}}^{-1}$ the steady state is reached. In
Fig.~\ref{fig:cw_energy} we have plotted the emitted phonon energy
${\mathcal{E}}_{\textrm{LA}}$ as a function of the Rabi frequency (red
crosses). We find that the dependence of the emitted energy on the Rabi
frequency is very well described by a linear relationship
${\mathcal{E}}_{\textrm{LA}}=\frac{1}{2} \hbar \Omega_{\textrm{R}} -
\frac{1}{8} \hbar \omega_{\textrm{pol}}$ shown as red line.

To understand this behavior let us revisit the model given by
Eq.~\eqref{eq:H} in the case of a driving with a cw light field. It is
instructive to rewrite the Hamiltonian in a basis rotating with the light
field. This leaves the phonon part and the exciton-phonon interaction
unchanged, while the exciton part including the exciton-light interaction
reads
\begin{equation}
\hat{H}_{\textrm{X}} = \hbar\left(\omega_x - \omega_{\textrm{L}}\right) |x\rangle\langle x|-
\frac{\hbar}{2}\Omega_{\textrm{R}} \left(|x\rangle \langle g| +
|g\rangle \langle x|\right) .  \label{eq:H_rot}
\end{equation}
In this basis the Hamiltonian is now time-independent thus implying energy
conservation, i.e., $\langle \hat{H} \rangle = \textrm{const}$. Since before
the light field is switched on the QD is in its ground state and no phonons
are present, the initial energy is zero. Thus the total energy should remain
zero. Introducing the exciton energy
\begin{equation}
E_{\textrm{X}} = \langle \hat{H}_{\textrm{X}} \rangle ,  \label{eq:E_X}
\end{equation}
this leads to
\begin{equation}
\langle \hat{H} \rangle = E_{\textrm{X}}+E_{\textrm{LA}}+E_{\textrm{X-LA}}=0 .
\label{eq:E_sum}
\end{equation}
In our simulations we calculate separately all three contributions to the
energy. Their sum is plotted by the small black diamonds in
Fig.~\ref{fig:cw_energy} which demonstrates that Eq.~\eqref{eq:E_sum} is
indeed very well satisfied thus confirming the consistency of the correlation
expansion approach and the accuracy of the numerics.

In addition to the emitted energy, Fig.~\ref{fig:cw_energy} shows the
absolute value of the exciton energy $|E_{\textrm{X}}|=-E_{\textrm{X}}$ (blue
squares) and one half of the interaction energy $E_{\textrm{X-LA}}$ (green
diagonal crosses). The black circles in Fig.~\ref{fig:cw_energy} denote the
difference between the (negative) exciton energy $|E_{\textrm{X}}|$ and
emitted phonon energy ${\mathcal{E}}_{\textrm{LA}}$, which obviously agrees
well with one half of the interaction energy. From these results we can
conclude that the emitted phonon energy consists of two parts: first the
energy $-E_{\textrm{X}}$ which is released by the relaxation of the exciton
and second an energy which is released by the formation of the polaron. Like
in the case of pulsed excitation this latter is given by one half of the
interaction energy $E_{\textrm{X-LA}}$. Furthermore we find that the exciton
energy is well described by the dependence $|E_{\textrm{X}}|=\frac{1}{2}
\hbar \Omega_{\textrm{R}} - \frac{1}{2} \hbar \omega_{\textrm{pol}}$ (blue
dashed line) while half of the interaction energy is well represented by
$\frac{1}{2}E_{\textrm{X-LA}}=-\frac{3}{8} \hbar \omega_{\textrm{pol}}$
(green dashed line). However the Figure also indicates that although there is
a very good agreement, these dependencies are most probably not exact. In
some ranges of Rabi frequencies there are small deviations of the order of at
most 10 percent of the polaron energy.

This rises the question whether at least the general behavior of the energies
shown in Fig.~\ref{fig:cw_energy} can be understood from the model.  For this
purpose the exciton Hamiltonian in the rotating basis given by
Eq.~\eqref{eq:H_rot} is diagonalized leading to the well known dressed states
picture. In our case the excitation is in resonance with the polaron shifted
exciton transition, thus with respect to the bare exciton energy there is a
detuning $\Delta=\omega_{\textrm{L}} - \omega_x=-\omega_{\textrm{pol}}$. In
the weak coupling limit, i.e., neglecting terms of second order in
$\omega_{\textrm{pol}}/ \Omega_{\textrm{R}}$ the energies of the dressed
states read
\begin{equation}
E_\pm  =  \frac{1}{2}\hbar\left( \omega_{\textrm{pol}}\pm  \Omega_{\textrm{R}} \right).
\end{equation}
Before the light field is switched on the QD is in its ground state
$|g\rangle$ with the exciton energy $E_{\textrm{X}}=0$. In the dressed state
picture this state translates into a superposition between the dressed states
$|\psi_{\pm}\rangle$ with the corresponding energies $E_\pm$. In this picture
the exciton-phonon interaction is given by
\begin{eqnarray}
\hat{H}_{\textrm{X-LA}} &=& \sum_{\mathbf{q}}\frac{\hbar}{2}\left(g_{\mathbf{q}}\hat{b}_{\mathbf{q}} +
g_{\mathbf{q}}^\ast \hat{b}_{\mathbf{q}}^\dag\right)\Bigl[ |\psi_{+}\rangle \langle \psi_{+}|
+|\psi_{-}\rangle \langle \psi_{-}| \nonumber \\
&& - |\psi_{+}\rangle \langle \psi_{-}|
-|\psi_{-}\rangle \langle \psi_{+}| \Bigr] ,
\label{eq:cp-dressed}
\end{eqnarray}
where again terms of higher order in the polaron energy have been neglected.
The interaction now has both diagonal and off-diagonal terms thus leading at
low temperatures to a relaxation from the upper to the lower dressed
state.\cite{glassl2011lon,luker2012inf,mccutcheon2010qua} At sufficiently
long times therefore the exciton system ends up in the lower dressed state
with the energy $E_{\textrm{X}}=E_{-}  = \frac{1}{2}\hbar\left(
\omega_{\textrm{pol}} - \Omega_{\textrm{R}}\right)$. Indeed, this agrees with
the dashed blue line in Fig.~\ref{fig:cw_energy} and is thus in excellent
agreement with the exciton energies found from the simulations.

The value $E_{\textrm{X-LA}} \approx -\frac{3}{4} \hbar
\omega_{\textrm{pol}}$ cannot so easily be understood. In the case of pulsed
excitation we have found $E_{\textrm{X-LA}} = -2f \hbar
\omega_{\textrm{pol}}$ (see Fig.~\ref{fig:E_tau}), which in the present case
would mean $E_{\textrm{X-LA}} = - \hbar \omega_{\textrm{pol}}$ since the
exciton occupation in the lower dressed states is $f=\frac{1}{2}$. On the
other hand, the dressed states interaction Hamiltonian in
Eq.~\eqref{eq:cp-dressed} has in its diagonal a coupling constant of
$g_{\mathbf{q}}/2$ which for the lower dressed state should lead to a polaron
energy reduced by a factor of 4 and thus to $E_{\textrm{X-LA}} = -\frac{1}{2}
\hbar \omega_{\textrm{pol}}$. The result found from our calculations is in
between these two values. This is in line with the results for the long-time
exciton state in the presence of a cw field discussed in
Ref.~\onlinecite{glassl2011lon}, where it has been shown that in particular
at low temperatures there are slight deviations from both a thermal
occupation of the dressed states and from calculations within a weak coupling
theory.

Nevertheless it should be pointed out that the discrepancies between the
models are in the range of fractions of the polaron energy. Typical Rabi
energies are much larger than the polaron energy, thus, as seen in
Fig.~\ref{fig:cw_energy}, the total energy transported away from the QD by
the LA phonons is in good approximation given by
${\mathcal{E}}_{\textrm{LA}}=\frac{1}{2} \hbar \Omega_{\textrm{R}}$.

\section{Conclusions}\label{sec:conclusions}

In conclusion, we have analyzed the spatiotemporal dynamics of the lattice
displacement in and around a QD driven by an external electromagnetic field.
In general, the displacement field consists of two parts, a localized one
which resides in the region of the QD and forms the acoustic polaron and an
outgoing strain wave consisting of a single or a series of wave packets. The
focus of our investigations has been on the one hand on the energy that is
transported by these phonon wave packets and on the other hand on the
coherence properties of the generated phonons. In the case of excitation by
Gaussian laser pulses we have found a non-trivial dependence of the emitted
acoustic energy on the pulse area and pulse duration. The strongest phonon
emission has been found when the time scale of the exciton dynamics,
determined by the instantaneous Rabi oscillation period, matches the
characteristic phonon time given by the oscillation period of the most
strongly coupled phonons. In this case each Rabi cycle of the exciton gives
rise to the emission of a pair of phonon wave packets, one with negative and
one with positive amplitude. If the exciton dynamics is either much slower or
much faster than the phonon dynamics the intensity of the wave packets
decreases and, in the limit of very short pulses, the energy may even be
further reduced by destructive interference of subsequent wave packets. While
for excitation by very short pulses with pulse areas given by odd multiples
of $\pi$ the emitted phonons are almost perfectly coherent, for longer pulses
an increasing contribution due to incoherent phonons shows up, as is
reflected by the difference between the total and the coherent acoustic
Poynting vector.

In the case of continuous excitations switched on instantaneously a series of
wave packets with decreasing amplitudes is emitted. The strongest damping is
again found for a resonant coupling between exciton and phonon system. Both
for much larger and much smaller Rabi frequencies the two systems become
increasingly decoupled. While the reduction of the damping of the Rabi
oscillations is almost symmetric with respect to the resonant Rabi frequency,
the acoustic Poynting vector is much larger for excitation with a Rabi
frequency above the phonon resonance because of the larger velocities of the
lattice atoms. The energy of the coherent phonons is emitted as a sequence of
pulses with maxima at the edges of the displacement wave packets. In
contrast, the total emitted energy is typically a rather smooth function of
time given by the envelope over the coherent pulses. Therefore, in a good
approximation half of the emitted energy is carried by coherent phonons and
the other half by incoherent phonons. The total energy which is transported
away from the QD by the phonons increases linearly with the Rabi frequency;
it is essentially given by one half of the Rabi energy $\hbar
\Omega_{\textrm{R}}$ which results mainly from the relaxation of the exciton
into the lower dressed state but has also a contribution from the polaron
dressing process.

Our results thus clearly demonstrate the close connection of the dephasing of
the excitonic degrees of freedom with the irreversibility caused by the
energy transport away from the QD region by the generated phonons. In
addition they shed new light on the coherence properties of these phonons
showing that the emitted phonons are in general not completely coherent, but
that the coherent phonons carry a rather large part of the total energy
(typically about 50 \%), which is different from other systems like biased
quantum wells, where even under optimal conditions the energy associated with
coherent phonons is orders of magnitude smaller than the energy transferred
to incoherent phonons.\cite{papenkort2010res}

\subsection*{Acknowledgments}

VMA wishes to thank the Deutsche Forschungsgemeinschaft for financial support
within the grant No.~AX 17/7-1.



\end{document}